\documentclass[mathleft
]{an}
\usepackage{graphicx}
\usepackage{times}
\usepackage{bm}
\usepackage{natbib}
\begin{document}

\Pagespan{789}{}
\Yearpublication{2007}%
\Yearsubmission{2007}%
\Month{11}%
\Volume{999}%
\Issue{88}%

\title{Zeeman-Doppler Imaging of late-type stars -- The surface magnetic field of II Peg}

\author{T.A. Carroll  \thanks{\email{tcarroll@aip.de}}, M. Kopf , I. Ilyin  \& K.G. Strassmeier 
}
\titlerunning{Zeeman-Doppler Imaging of II Peg}
\authorrunning{T.A. Carroll et al.}
\institute{
Astrophysikalisches Institut Potsdam, An der Sternwarte 16, 
D-14482 Potsdam, Germany
}

\received{11 Nov 2007}
\accepted{22 Nov 2007}
\publonline{later}


\abstract{%
  Late-type stars in general possess complicated 
  magnetic surface fields which makes their detection and in particular their
  modeling and reconstruction challenging. In this work we present a new Zeeman-Doppler
  imaging code which is especially designed for the application to late-type stars. 
  This code uses a new multi-line cross-correlation technique
  by means of a principal component analysis to extract and enhance the quality of individual 
  polarized line profiles.
  It implements the full polarized radiative transfer equation
  and uses an inversion strategy that can incorporate prior knowledge based on solar analogies.
  Moreover, our code utilizes a new regularization scheme which is based on local maximum entropy 
  to allow a more appropriate reproduction of complex surface fields as those expected for late-type stars.
  In a first application we present Zeeman-Doppler images of II Pegasi
  which reveal a surprisingly large scale surface structure with one predominant (unipolar) 
  magnetic longitude which is mainly radially oriented. }

\maketitle
\section{Introduction}

Even though the technique for Zeeman-Doppler imaging was introduced almost two decades ago
by \citet{Semel89} there are only a very small number of magnetic surface images for late-type stars
\citep{Donati99,Petit06}. The reason for this fact is not so much the
limited number of available targets than the inherent complexity of the surface magnetism
which may lead to a severe reduction or even the mutual cancellation 
of circular polarization profiles (Stokes $V$) by different magnetic polarities. 
The linear polarization profiles (Stokes $Q$ and $U$)
which are, as a rule of thumb, at least one order of magnitude lower than the circular polarization
are even harder to detect.
Cross-correlation techniques, in its simplest form the multi-line technique of \citet{Semel96} or
the more elaborate technique of Least-Square Deconvolution (LSD) \citep{Donati97a} are used
to extract and boost the polarization signal to a level were a detailed diagnostic
of surface magnetic fields become feasible. But the trade off is the reduced
interpretability of the retrieved \emph{mean} line profile.

In an effort to provide the necessary analysis and diagnostic tools for the next generation 
of spectropolarimeter (PEPSI) at the 8.4 m Large Binocular Telescope (LBT) \citep{Strass03,Strass07},
which, for the first time, provide the instrumental capabilities to detect and measure 
the magnetic fields of a significant number of active late-type stars, 
we have developed a novel multi-line reconstruction technique and a full Stokes 
Zeeman-Doppler imaging code. 
In the following we give a short presentation of the new line profile reconstruction technique
as well as of our new ZDI code \emph{IMap}. In a first application we also present magnetic field 
surface maps of II Pegasi.

\section{The Zeeman-Doppler Imaging code}

The forward module of the ZDI code \emph{IMap} 
incorporates the full numerical solution of the polarized radiative transfer equation
by means of the so called Diagonal Lambda Operator method (DELO) \citep{Rees89}.
The code works on the basis of precalculated and/or interpolated Kurucz model atmospheres \citep{Kurucz93}.
Parameters for the individual spectral lines are retrieved from the Vald database \citep{Piskunov95}
and line blends are full accounted for. The surface is parameterized on a variable equal-area or
equal-degree partition with a minimum area of 1 $\times$ 1 degree. For each surface element a local
Stokes vector is calculated with respect to its position and atmospheric parameters
(Doppler velocity, bulk velocity, temperature, pressure, 
LOS magnetic field, micro- and macroturbulence). The atmospheric
parameters are also allowed to vary along their local vertical 
direction which facilitates the complete description of a 
3-dimensional atmosphere. Center-to-limb variations in the
model atmosphere as well in the atmospheric parameters are fully account for by adjusting
the depth stratification for each surface element with respect to the local reference frame of the observer.
Field structures (in temperature and magnetic fields) can either be described by setting individual
surface elements or using spherical harmonics.

The inverse module which is responsible for the fitting of the observed profiles by adjusting
the model parameters implements two optimization algorithms. A Levenberg-Marquardt method and
a conjugate gradient method \citep{Press92}. Both algorithms are complemented by a regularization
functional. We have used a similar maximum entropy regularization as formulated by \citet{Brown91}.
which reads 
\begin{equation}
S = - \sum_{i} \left ( |P_i| + \alpha \right ) \log \frac{|P_i| + \alpha}{|m_i| + \alpha} - 1 \; ,
\label{entropy1}
\end{equation}
where the index $i$ runs over all surface elements, $\alpha$ is a small positive value to ensure that the 
entropy function is always greater than zero and $m_i$ is the so called default image of the surface element
or segment. 
It is this default value which can give the entropy function a
quite different characteristic. Even though the importance of the default images was never described in 
great detail in the ZDI literature the default images of the elements are usually set all to an equal value 
which is in most cases the global average of the parameter under consideration (i.e. magnetic field strength, 
inclination, azimuth or temperature). 
This is in fact problematic becuase this constrain is invariant under random mixing of the surface 
elements (pixels) and therefore inadequate for small scale (spatial) variations with great local 
differences in the parameters but also for smooth large scale variations in the parameters 
\citep{Brown91,Donati97b,Piskunov02}. We therefore propose an 
adaptive \emph{local} entropy function 
where the default images $m_i$ are retrieved by the actual mean value of the surrounding 
neighborhood pixels such that
\begin{eqnarray}
S = - \sum_{i} \left ( |P_i| + \alpha \right ) \hspace{3.6cm} \nonumber \\ 
 \times \log \frac{|P_i| + \alpha}{\frac{1}{k}\sum_{j=0}^{k}\beta(j) |m_j| + \alpha} - 1 \;.
\label{entropy2}
\end{eqnarray}
Here the sum in the logarithm runs over the entire neighborhood of k elements and 
$\beta(j)$ is an extra term which includes a particular positve weighting.
As the regularization term in the penalty function becomes important (compared to the regular squared error function) 
as soon as the fit of the individual Stokes profiles reaches the noise level of the observations, 
the default image is also the appropriate place to incorporate 
prior knowledge or assumptions that are not purely data driven (determined by the local gradient of the error function).
We will described this topic in greater detail in a forthcoming paper.

The ZDI code can simultaneously or in a consecutive manner retrieve the temperature (Doppler imaging) and the magnetic field
distribution (Zeeman Doppler imaging) by using the complete Stokes vector ($I,U,Q,V$) or optionally 
only the Stokes $I$ and $V$ component.

\section{Multi-Line principal component reconstruction}

Most of the measured spectropolarimetric signals and profiles of individual spectral lines do not allow 
an in-depth analysis and interpretation which led to the development of 
powerful multi-line techniques such as LSD \citep{Donati97a}.
The success to which this method is able to extract and boost the signal quality comes
at the expense of using approximate assumptions (weak fields, a-priori model atmospheres and Gaussian line profiles)
and, moreover, of loosing the ability to use single spectral line profiles. 
Only a kind of \emph{mean} line profile can be extracted by this technique which therefore has to be described
in terms of  artificial mean line parameters (e.g. excitation potential, Lande factor, oscillator strength etc.). 
In order to avoid simplified a-priori assumptions and making use of single spectral line profiles
but benefiting at the same time from a multi-line cross-correlation technique 
we propose a multi-line principal component reconstruction technique.
The basic idea is to use the redundancy and the variance in the observed data (the Stokes profiles) 
to determine a new coordinate system where each new coordinate axis accounts for 
a maximum of the variance in the original data.
This can be conveniently expressed in terms of the covariance matrix $\bm{C_x}$
of the individual observed profile vectors $\bm{x_i}$, which reads in the velocity domain
\begin{equation}
\bm{C_x} = \sum_n \left ( \bm{x}_n(v) - \bar{\bm{x}}(v) \right ) \left ( \bm{x}_n(v) - \bar{\bm{x}}(v) \right )^T \; ,
\label{covar1}
\end{equation}
where $v = c \Delta\lambda/\lambda$ and $n$ is the number of individual spectral line profiles used for
the analysis 
and $\bar{\bm{x}}$ is the mean Stokes profile of all spectral lines.
The new set of coordinate axes which accounts for the maximum variance in the observed data (Stokes spectra)
can then be determined by calculating the eigenvectors of the covariance matrix Eq. (\ref{covar1}). This is exactly
what the so called Principal Component Analysis (PCA) or Karhunen-Loeve transformation accomplishes \citep{Bishop95}. 
We use the PCA method to decompose the entire set of observed Stokes spectra into a new coordinate
system. This procedure will project the most coherent and systematic features in the observed Stokes profiles 
into the first few eigenvectors with the largest eigenvalues while the uncoherent features (i.e. noise) will 
be mapped to the less significant eigenvectors (with low eigenvalues). 
Since the covariance matrix Eq. (\ref{covar1}) is symmetric it results in a set of orthogonal eigenvectors 
which we can used to decompose all observed Stokes spectra (with out any loss) into the new basis of 
eigenvectors $\bm{u_l}$, as 
\begin{equation}
\bm{x}_k(v) = \sum_l \alpha_{k,l} \bm{u}_l(v)  \; ,
\label{project}
\end{equation}
where $\alpha_{k,l} = \bm{x}_k(v) \bm{u}_l(v)$ is the scalar product (the projection or the cross-correlation)
between the observed Stokes profile $\bm{x}_k(v)$ and the eigenvector $\bm{u}_l(v)$. 
If we make use of the well know decomposition and dimensionality
reduction capabilities of the PCA method \citep{Bishop95} and using only the first few eigenvectors
to reconstruct the original spectra $\bm{x_k}$ such that
\begin{equation}
\bm{x}_k(v) = \sum_{m < l} \alpha_{k,m} \bm{u}_m(v)  \; ,
\label{reconpro}
\end{equation}
we can reconstruct the individual Stokes profiles 
to an extent that the majority of the characteristic features of one particular line are well 
reproduced with a minimum of uncorrelated effects. 
This allows us to use a large set of observed Zeeman-sensitive spectral line profiles   
to get rid of the uncoherent effects like noise while retaining the 
common and systematic features (in all observed line profiles) which are induced 
by the Doppler- and Zeeman-effect. 
Since this method allows the reconstruction of a single line profile, all the known line parameter
can be used in the following DI and ZDI process.
\begin{figure}
\includegraphics[width=41mm,height=36mm]{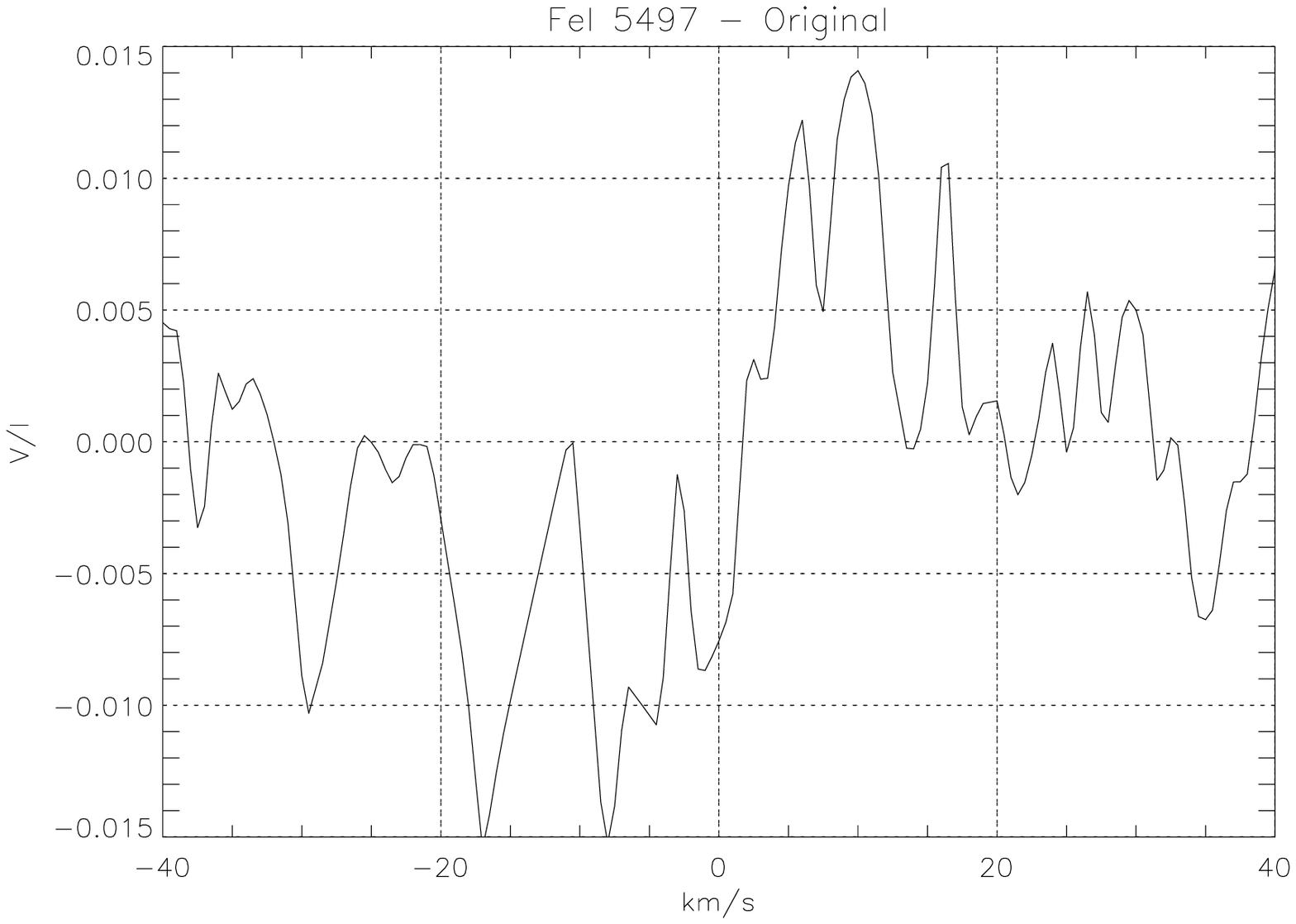}
\includegraphics[width=41mm,height=36mm]{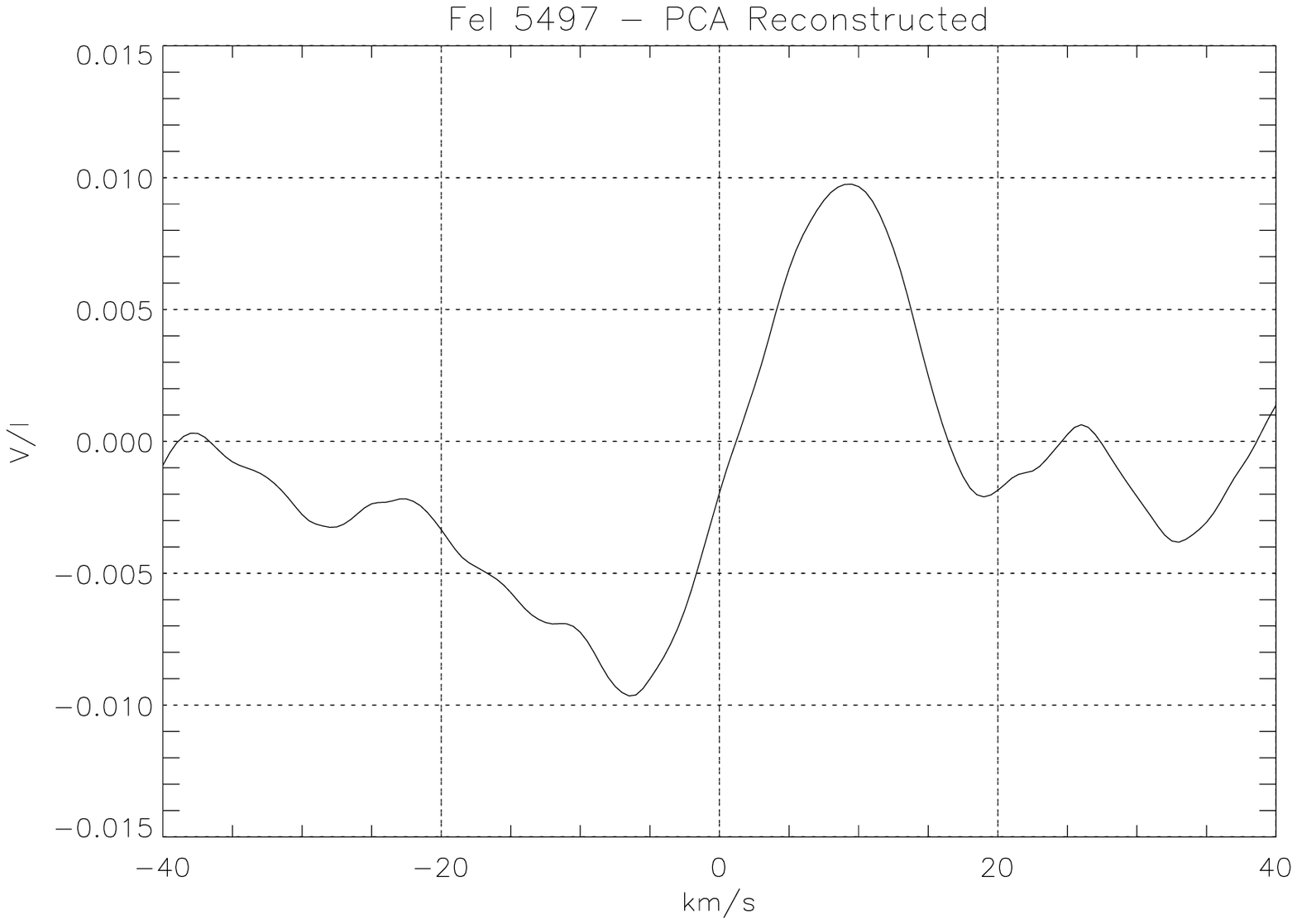}
\caption{On the left the original observed Stokes $V$ profile of the FeI 5497 \AA\ line and on the
right side the multi-line PCA reconstructed Stokes $V$ profile for one rotational phase.}
\label{Fig:origProf}
\end{figure}
In Fig. \ref{Fig:origProf} we show a Stokes $V$ line profile for the iron line FeI 5497 \AA\ which has been observed for
II Peg with the SOFIN spectrograph at the Nordic Optical Telescope (NOT) \citep{Tuominen99}. 
For the multi-line PCA reconstruction we have used 18 magnetic sensitive spectral lines (\ion{Fe}{i},\ion{Ca}{i} ,\ion{C}{i}) 
in a wavelength range
between 4600 and 6600 \AA. To reconstruct the individual line profile we have used the three largest principal components.
The Stokes $V$ line profile -- although only 18 spectral lines in the covariance matrix are used -- 
exhibit a much smoother and less noisy behavior and is due to the PCA 
analysis more reliable in terms of its interpretability.
This method will be fully described and statistically analyzed in a forthcoming paper. 

\section{The surface magnetic field of II Peg}
\begin{figure}[h]
\includegraphics[width=70mm,height=70mm]{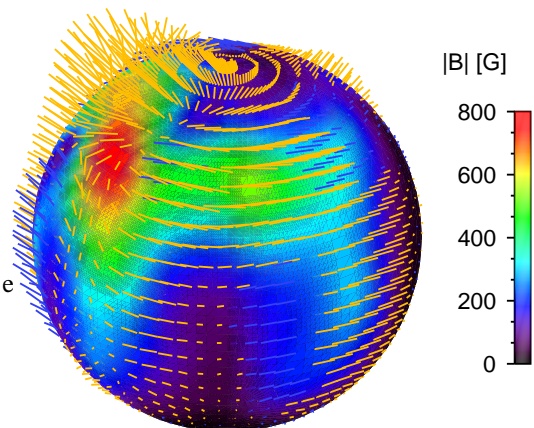}
\caption{Surface magnetic field of II Peg. The thin white lines indicating the field vector 
for each surface pixel and the underlying color gives the absolute field strength.}
\label{Fig:surfIIPeg}
\end{figure}

\begin{figure}
\includegraphics[width=80mm,height=40mm]{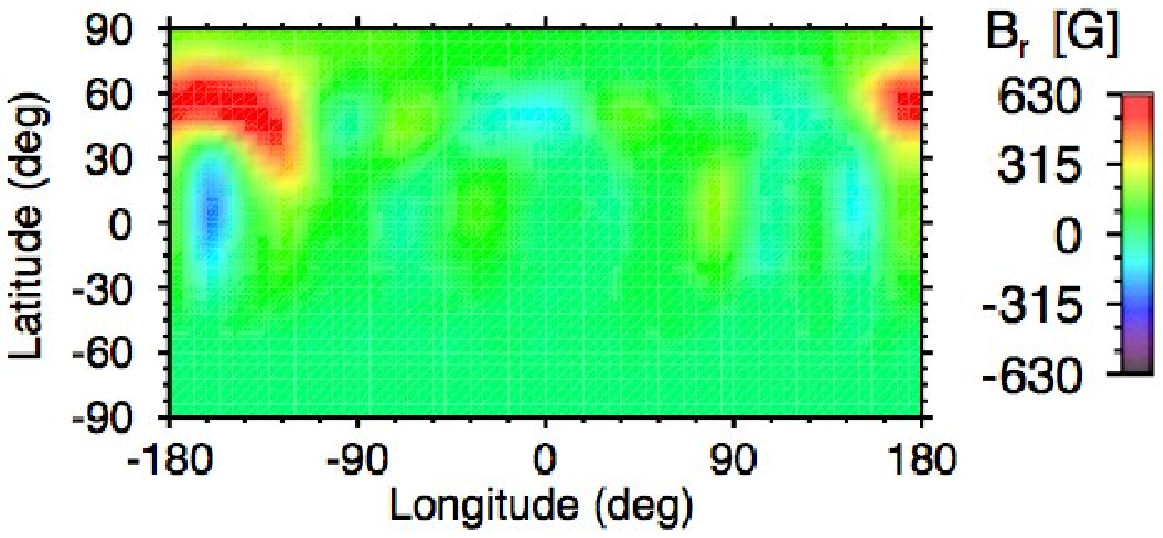}
\includegraphics[width=80mm,height=40mm]{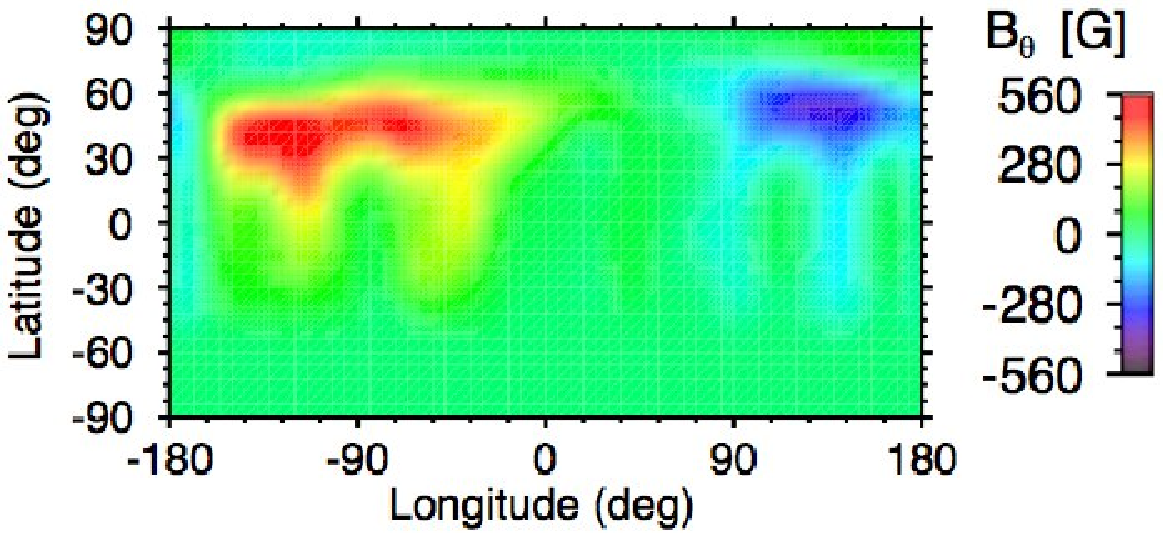}
\includegraphics[width=80mm,height=40mm]{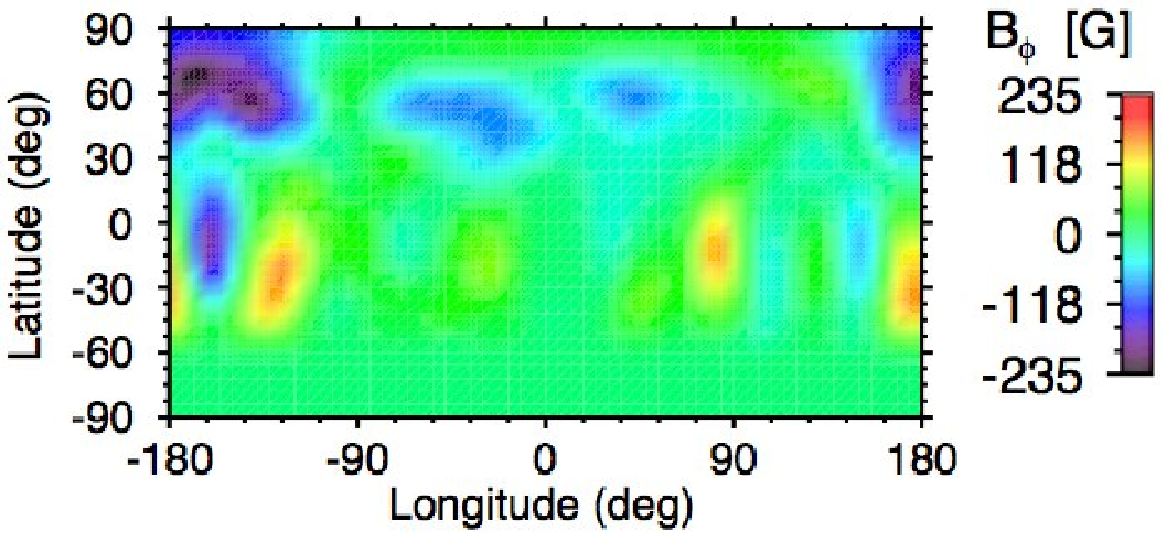}
\caption{Mercator plots of the surface magnetic field of II Peg. Top: the radial magnetic field; Middle: 
the azimuthal magnetic field component; Bottom : The meridional magnetic field component.}
\label{Fig:mercIIPeg}
\end{figure}

In a first application we have applied our code to spectropolarimetric observation of the K1 subgiant star II Peg
observed with the SOFIN spectrograph at the NOT.
The data (Stokes $I$ and Stokes $V$ ) were collected in an observing campaign in 2004 and cover a whole 
rotational period. The stellar parameters for the inversion are taken from \citet{Berdyu98}.
\begin{figure}
\includegraphics[width=70mm,height=80mm]{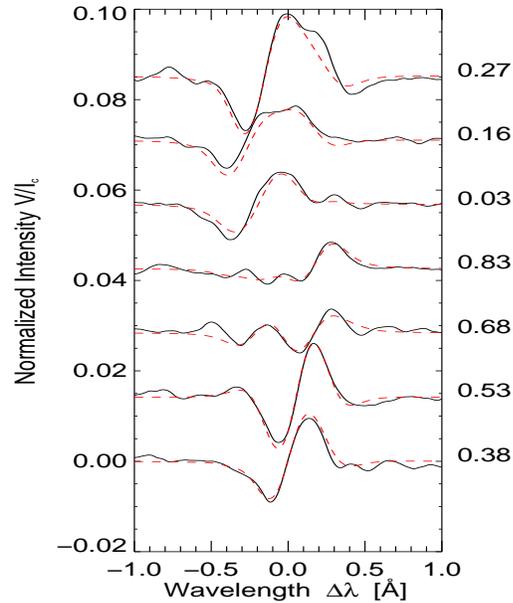}
\caption{Stokes $V$ profile fits (dashed) to the observed profiles (solid) of the iron line FeI 5497 \AA\ for 
different rotational phases.}
\label{Fig:allPhases}
\end{figure}
Prior to the actual Zeeman-Doppler inversion we have used our code in DI mode to retrieve the temperature distribution
of the surface of II Peg, where we have used the CaI 6439 \AA\ line. 
This information was subsequently used in the ZDI inversion to determine the magnetic field from the phase resolved
Stokes $V$ profiles of the FeI 5497 \AA\ line, which was reconstructed by the multi-line 
PCA technique described in the preceeding section.
After careful tests with different initialization and setups 
the observed profiles are well reproduced (see Fig.\ref{Fig:allPhases}).
Despite the fact we have only used Stokes $V$ profiles the surface exhibits a rich magnetic surface 
structure as can be seen in Fig. \ref{Fig:surfIIPeg} and Fig. \ref{Fig:mercIIPeg}. One of the 
conspicuous features in the ZDI map
is that the magnetic activity is mainly located in one active longitude 
where the magnetic field shows a strong radial component. 
It will be interesting
to see in forthcoming studies which expands over longer time periods if and in which way these 
regions of magnetic activity are associated with the active longitudes
in the temperature distribution found by \citet{Berdyu99}.
As has been observed in former ZDI approaches \citep{Donati99} we also
notice only a small correlation between the dark and cool starspots as retrieved by our
DI run and the magnetic field distribution from the ZDI.
This seems to be plausible because of the strong suppression of the photon flux in the
spot regions which are between 700 and 1000 K cooler than the quiet and hot parts of the surface.
It should be mentioned that the retrieved surface field exhibit a strong imbalance in the magnetic polarity and we
believe that this imbalance of magnetic flux results to a large extent from the inability of the circular polarization 
signal to capture the whole magnetic flux of the star, e.g. the probably strong magnetic fields in the spot 
regions of II Peg seem to be virtually absent in the Stokes $V$ profiles.

\subsection{Summary and conclusion}
We have presented a new Zeeman-Doppler imaging code which is particularly designed for the demands
of retrieving surface temperature and magnetic field vector distributions of active  
late-type stars.
Moreover, we have introduced a novel multi-line PCA reconstruction technique which relies on a minimal
number of a-priori assumption and allows to extract and boost the signal-to-noise ratio of
individual spectral line profiles.
A first application shows the potential of the ZDI code and the new multi-line technique.
Even though the field structure results in a smooth distribution it should be noted that the 
reconstruction of the surface magnetic field is far from being a well posed problem.
Small differences in the underlying initialization, in the regularization function or 
the temperature distribution as well as changes in the optimization strategy can lead to 
rather drastic effects in the recovered magnetic surface fields (see forthcoming paper of 
Carroll et al. 2008)). 
It is therefore of particular interest to make the linear 
polarization profiles available and to retrieve circular polarized profiles with 
the best possible accuracy and signal-to-noise ratio  
to provide more constrains to the solution of the ZDI inversion and to reveal more 
of the hidden magnetic field information in the spectra.

\end{document}